# Dynamics of anisotropic oxygen-ion migration in strained cobaltites

*Qinghua Zhang, ‡, § † Fanqi Meng, ‡,† Ang Gao, ‡,∥ Xinyan Li, ‡,∥ Qiao Jin, ‡,∥ Shan Lin, ‡,⊥ Shengru Chen, ‡,∥ Tongtong Shang, ‡,∥ Xing Zhang, # Haizhong Guo, ▽ Can Wang, ‡,○ Kui-juan Jin, ‡,∥,○ Xuefeng Wang, ‡,○ Dong Su, ‡,○ Lin Gu‡,∥,○,\* and Er-Jia Guo‡,⊥,○,\**

‡ Beijing National Laboratory for Condensed Matter Physics, Institute of Physics, Chinese Academy of Sciences, Beijing 100190, China
§ Yangtze River Delta Physics Research Center Co. Ltd., Liyang 213300, China
∥ School of Physical Sciences, University of Chinese Academy of Sciences, Beijing 100049, China
⊥ Center of Materials Science and Optoelectronics Engineering, University of Chinese Academy of Sciences, Beijing 100049, China
# Beijing National Laboratory for Molecular Sciences, Key Laboratory of Molecular Nanostructure and Nanotechnology, Institute of Chemistry, Chinese Academy of Sciences, Beijing 100190, China.
▽ School of Physics and Microelectronics, Zhengzhou University, Zhengzhou 450001, China
○ Songshan Lake Materials Laboratory, Dongguan, Guangdong 523808, China

E-mail: ejguo@iphy.ac.cn; l.gu@iphy.ac.cn
† These authors contribute equally to this work.

**ABSTRACT**

Orientation control of oxygen vacancy channel (OVC) is a highly desirable for tailoring oxygen diffusion as it serves fast transport channel in ion conductors, which is widespread exploited in solid-state fuel cells, catalysts, and ion-batteries. Direct observation of oxygen-ions hopping towards preferential vacant sites is a key to clarifying migration pathways. Here we report the anisotropic oxygen-ion migration mediated by strain in ultrathin cobaltites via in-situ thermal activation in an atomic-resolved transmission electron microscopy. Oxygen migration pathways are constructed on the basis of the atomic structure during the OVC switching, which is manifested as the vertical-to-horizontal OVC switching under tensile strain, but the horizontal-to-diagonal switching under compression. We evaluate the topotactic structural changes to OVC, determine the crucial role of tolerance factor for OVC stability and establish the strain-dependent phase diagram. Our work provides a practical guide for engineering OVC orientation that is applicable ionic-oxide electronics.

**TEXT**

Oxygen-ion migration in transition metal oxides has attracted an increasing interest in fundamental research and applications in solid oxide fuel cells,[1-3] fast catalytic reaction,[4] resistive switching memories,[5] electrochemical sensing,[6] and *etc*. The oxygen-ion migration in solids is ascribed to the hopping towards lower energetic vacant oxygen sites. Ordering of the oxygen vacancy ($V_{\ddot{o}}$) emerges when their concentration reaches a fractional threshold in oxide parents,[7] leads to the formation of the oxygen vacancy channel (OVC)[8-11] and stimulates various emergent quantum phenomena.[12-18] More importantly, the creation or reorientation of OVC introduce topotactic phase transition such as the transformation from perovskite to brownmillerite (BM), resulting in dramatic change in physical and chemical properties including the metal−insulator transition, magnetic phase transition, and anisotropic ion conduction.[11,19-24] The OVC usually serves as a fast channel for oxygen vacancy diffusion,[17,18] play the key role in regulating migration pathway of oxygen ions and dominating the dynamic response to external stimulus. Thus, controllable orientation of OVC is desirable for tailoring oxygen diffusion as its widespread exploited in solid-state fuel cells, catalysts, and ion-batteries. A thoughtful understanding of topochemical aspects of the oxygen-ion migration will give us further insight into the control parameters of oxygen-ion migration pathways and the utilization of OVC in designing solid oxygen-ion conductors.[25-27]

Strain engineering is a universal strategy to tune the atomic configuration of OVCs by directly altering the cation-oxygen bond length and oxygen vacancy formation energy ($E_{V\ddot{o}}$). Moving an oxygen ion out of its equilibrium position along a specific migration pathway to the neighboring vacant lattice site is determined by the strain state of epitaxial thin films. It is reported that the orientation of OVCs in cobaltites,[28] chromates,[29] and ferrites[30] thin films can be readily controlled by strain. For instance, the periodicity and orientation of $V_{\ddot{o}}$ layers are parallel (perpendicular) to the interface when the films under compressive (tensile) strain.[31]

Recently, an unexpected zigzag-like $V_{\ddot{o}}$ ordering in a compressively strained LaCoO$_{2.5}$ film has been found, whereas the unique structure cannot be formed through annealing a tensile-strained LaCoO$_3$(LCO$_3$) film.[32] Previously, the OVC formation under tensile strain has been studied by high-dose electron beam irradiation.[33,34] Dynamic process in differently strained LaCoO$_x$ (LCO$_x$) films towards further oxygen reduction to the metastable phases has not been visualized directly so far.

Here we report an *operando* transmission electron microscopy (TEM) investigation of dynamic structural evolution in LCO$_x$ films during the thermally annealing. We observe that the final-stabilized LCO$_{2.5}$ phases are distinct when the pristine LCO$_3$ films under compressive and tensile strain states. The OVC evolutions of strained LCO$_x$ films are recorded as a function of both annealing temperature and annealing time. By imaging the atomic-scale boundaries of OVC switching, this work provides the direct evidence of strain-mediated $V_{\ddot{o}}$ migration pathway from an octahedron. The associated lattice distortion and its strain dependency are quantified theoretically by mapping out the OVC phase diagram at the atomic scale.

**Strain dependence of OVCs' orientation in ultrathin cobaltites**

Previous theoretical and experimental work[35-40] demonstrate that $E_{V_{\ddot{o}}}$ in perovskite-type oxides depends on the strain and is highly anisotropic. Tensile strain dramatically reduces $E_{V_{\ddot{o}}}$, whereas the compressive strain only changes $E_{V_{\ddot{o}}}$ slightly. We choose LCO$_3$ as our protocol system due to the following two reasons. First of all, cobaltites have relatively low $E_{V_{\ddot{o}}}$ (~ 1.5 eV) and oxygen migration barrier (~ 0.8 eV), compared to other 3$d$ transition metal oxides.[40] Secondly, earlier work shows that the stoichiometric LCO$_3$ epitaxial films exhibit an unconventional strain relaxation behavior, resulting in the stripe-like domain patterns due to the formation of OVCs.[41-44] The pseudocubic (*pc*) lattice constant of bulk LCO$_3$ is $a_{pc}$ = 3.81 Å (**Figure 1**a). Single-crystalline SrTiO$_3$ ($a$ = 3.905 Å) and LaAlO$_3$ ($a_{pc}$ = 3.79 Å) substrates may introduce −0.5% (compressive) and +2.5% (tensile) strain to the as-grown LCO$_3$ films, respectively. Therefore, the LCO$_3$ thin film is expected to exhibit the unique structural evolution of OVC under different strain states.[45] We choose the LaAlO$_3$ as capping layer on LCO$_3$ films because it has a lower oxygen migration barrier (~ 0.63 eV) and higher oxygen vacancy formation energy (4.2-6.5 eV) than those of cobaltites.[46-48] Therefore, the oxygen could extract out of the films without deteriorating the crystal structure of the capping layer. X-ray diffraction and reflectivity measurements confirm the highly epitaxial, coherently growth, and smooth interface/surface in the as-grown LaAlO$_3$/LCO$_3$ bilayers (Figure S1). The absence of periodic dark-strips in the pristine LCO$_3$ films of both strain states indicate that all LCO$_3$ films are nearly stoichiometric (Figure S2). The oxygen had been gradually extracted from LCO$_3$ films by gradually increasing temperature at rate of 5 °C/min in the ultrahigh vacuum environment (~ 10$^{-9}$ Torr) of STEM (Figure S3 and S4). Figures 1b and 1c show the immediate high-angle annular dark-field (HAADF) STEM images of LCO$_{2.67}$ phases under tensile and compressive strain states at an annealing temperature of $T_a$ = 400 °C, respectively. As oxygen-ion escaped out of the LCO$_3$ films, the increased "*dark-stripes*" can be visualized in both samples. The LCO$_{2.67}$ phases contain alternatively stacked one CoO$_4$ tetrahedral and two CoO$_6$ octahedral layers.[41] As shown in Figure 1b, tensile strain would favor the creation of ordered $V_{\ddot{o}}$ in the films along the out-of-plane direction because the $V_{\ddot{o}}$-related chemical expansion can reduce the lattice misfit strain.[41-44] On the contrary, compressive strain leads to in-plane $V_{\ddot{o}}$ orderings, which is parallel to the interface due to the concomitant lattice elongation along the out-of-plane direction (Figure 1c).

Spontaneously, the ordered OVCs are formed within the tetrahedral CoO$_4$ layer in every one-third atomic plane,[49] producing the *dark-stripe* contrast in the HAADF images due to the increased La-La distance. Under tensile strain (on a SrTiO$_3$ substrate), $V_{\ddot{o}}$ sites prefer to the ordered arrangement along the out-of-plane direction, resulting in vertical $V_{\ddot{o}}$ stripes mostly with 3*a* periodicity (denoted as 3*a*-LCO$_{2.67}$ hereafter). In contrast, horizontal $V_{\ddot{o}}$ stripes with 3*c* periodicity (denoted as 3*c*-LCO$_{2.67}$ hereafter) appeared in the films grown on the LaAlO$_3$ substrates, corresponding to the $V_{\ddot{o}}$ sites aligned in the in-plane CoO$_4$ layers.

**Evolution of OVCs in strained LCO$_x$ films**

The orientation of OVCs switches from the out-of-plane to the in-plane orientation when a small amount of oxygen vacancies was extracted under tensile strain at a slightly higher temperature of 450 °C. As shown in **Figure 2**a, the horizontal $V_{\ddot{o}}$ stripe domains (3$c$-LCO$_{2.67}$) firstly appear randomly in the 3$a$-LCO$_{2.67}$ films and then propagate laterally. As increasing annealing time, the horizontal $V_{\ddot{o}}$ stripe domains emerge on the right side, and then moved towards the left region with vertical $V_{\ddot{o}}$ stripes within the tens of seconds (the minimum imaging time frame in scanning mode). It is hampered by an edge dislocation labeled by a blue triangle (Figure 2b). After annealing at $T_a$ = 450 °C around 2 minutes, the horizontal $V_{\ddot{o}}$ stripes crossed the edge dislocation and overspread the whole film. After ~ 4 minutes of high-temperature annealing, the 3$c$-LCO$_{2.67}$ film gradually transforms into the brownmillerite (BM) phase with 2$c$ periodicity (denoted as 2$c$-LCO$_{2.5}$) and finally stabilizes at 2$c$-LCO$_{2.5}$ phase after 6-minutes-annealling time (Figures 2c and 2d). Please note that the OVCs' alignment switches when the 3$a$-LCO$_{2.67}$ changes to 3$c$-LCO$_{2.67}$ without changing oxygen content. We could extract the moving speed of domain boundary is ~ 15 nm/mins. However, the domain boundaries change from vertical to horizontal alignment with further extracting oxygen to 2$c$-LCO$_{2.5}$.

The evolution behavior of OVCs in LCO$_x$ films under compressive strain is dramatically different to that of a film under tensile strain. At $T_a$ = 500 °C, we observe a new $V_{\ddot{o}}$ ordered phase with diagonal-aligned dark-stripes, which can be called n-LCO$_{2.5}$ as reported in our recent work.[32] The diagonal $V_{\ddot{o}}$ stripes firstly appear randomly in the 3$c$-LCO$_{2.67}$ films as indicated by the inclined purple arrows in Figure 2e. As increasing annealing time, the n-LCO$_{2.5}$ domains propagate laterally toward two sides in the moving speed of domain wall of ~ 7 nm/mins, accompanied by the disappearance of horizontal $V_{\ddot{o}}$ stripes (Figures 2f and 2g). After around 6 minutes, the 3$c$-LCO$_{2.67}$ completes the structural transformation to the $n$-LCO$_{2.5}$ phase in the entire film (Figure 2h). Both HAADF and annular bright field (ABF) images of n-LCO$_{2.5}$ were recorded for identifying the atomic positions of each element (Figure S5). From ABF images, we could visualize the tilt and distortion of CoO$_5$ square pyramids, which are dramatically distorted close to $V_{\ddot{o}}$ cites. The periodic OVCs along the diagonal direction were visualized directly, as indicated by bright contrast in ABF images. The dynamic process of dark stripe formation and evolution under thermal annealing exhibits completely different pathways depending on epitaxial strain.

**Atomic view of domain boundaries and strain distributions in LCO$_x$ films**

To reveal the atomic-resolved structural transition of OVCs, we compared the changes on the periodicity and orientation of $V_{\ddot{o}}$ stripes in the high-magnified HAADF images of LCO$_x$ films grown on SrTiO$_3$ and LaAlO$_3$ substrates, shown in **Figures 3**a and 3d, respectively. Figure 3c shows the domain boundary between 3$a$-LCO$_{2.67}$ and 3$c$-LCO$_{2.67}$. The atomic arrangement and structure model of domain boundary are shown in Figure 4c. We find that the vertical $V_{\ddot{o}}$ stripes are not connected directly to the horizontal ones but always blocked by an octahedral layer, marked in red shadow. The possible migration path of oxygen ions is illustrated in Figure S6. The oxygen ions will migrate from diagonal octahedron to the nearby tetrahedron, resulting in the rearrangement of dark stripes from vertically-aligned to horizontally-aligned. Meanwhile, the oxygen ions flow from an octahedron to a tetrahedron within the 3$a$-LCO$_{2.67}$ domains in order to keep an octahedral layer at the domain boundary. Macroscopically, the 3$c$-LCO$_{2.67}$ domains grow and 3$a$-LCO$_{2.67}$ domains vanish. Finally, the dark stripes change its periodicity from 3$c$ to 2$c$ by increasing numbers of $V_{\ddot{o}}$. Eventually, the 3$c$-LCO$_{2.67}$ undergoes a topotactic structural transformation into BM-phase 2$c$-LCO$_{2.5}$. Strain evolution at the domain boundary between 3$a$-LCO$_{2.67}$ and 3$c$-LCO$_{2.67}$ was also quantitatively analyzed by La-La distance mapping (Figure 3b and Figure S7). An obvious strain distribution can be visualized by the color contrast of lattice constants. A uniform contrast around domain boundary suggests that the lattice mismatches have been well accommodated by horizontal stripes through the orientation switching of OVCs, indicating a smooth transition from 3$a$-LCO$_{2.67}$ to 3$c$-LCO$_{2.67}$ domains.

Figure 3f shows the high-magnified HAADF image around the domain boundary between 3$c$-LCO$_{2.67}$ and n-LCO$_{2.5}$. The horizontal $V_{\ddot{o}}$ stripes with 2$c$ and 3$c$ periodicities coexist in LCO$_{2.67}$, indicating the oxygen content of LCO$_x$ films derivates from a stochiometric LCO$_{2.67}$ and approaches towards LCO$_{2.5}$. The horizontal $V_{\ddot{o}}$ stripes in 3$c$-LCO$_{2.67}$ with alternative [CoO$_6$] octahedron and [CoO$_4$] tetrahedron layers switch to the uniformed [CoO$_5$] square pyramids in n-LCO$_{2.5}$. There is no blocking layer between two distinct domains. The horizontal $V_{\ddot{o}}$ stripes are connected directly to the diagonal $V_{\ddot{o}}$ site of n-LCO$_{2.5}$ phase, indicating a comparable short oxygen-ion migration pathway. In this case, the oxygen ions migrate from the neighbored octahedron and tetrahedron vertically (Figure S8), resulting in the formation of pyramids. Different to the tensile strained films, the n-LCO$_{2.5}$ phase remain coherently strained by LaAlO$_3$ substrate (Figure 3e and Figure S9). Although the in-plane and out-of-plane lattice constants changes alternatively along the diagonal direction within the n-LCO$_{2.5}$ layer, the substrate's misfit strain can be perfectly transferred to the capping layers.

**Strain dependent oxygen migration and phase diagram of OVCs in cobaltites.**

To verify the influence of elastic strain on $E_{V\ddot{o}}$ in LaCoO$_3$ system, we performed the first-principles calculations based on density function theory (**Figures 4**a and 4b). $E_{V\ddot{o}}$ reduces monotonously as the in-plane stain changes from compressive to tensile strain state. Experimentally, we indeed observe a lower transition temperature of 3$a$-LCO$_{2.67}$ than that of 3$c$-LCO$_{2.67}$. More importantly, the strain-dependent $E_{V\ddot{o}}$ determines the OVC orientation when the $V_{\ddot{o}}$ stripes form at the early stage. The phase transition from 3$a$-LCO$_{2.67}$ to 3$c$-LCO$_{2.67}$ undergoes a OVC switching by 90 degrees. The oxygen ions rearrange within the films without changing oxygen content. Strain-dependent OVC in 3$a$-LCO$_{2.67}$ and 3$c$-LCO$_{2.67}$ is reasonable in terms of minimization of the elastic energy at the interface between films and substrates. We believe that the oxygen extracts out of the strained LCO$_3$ films may have different migration pathways as the OVCs serve as oxygen ion passes. To form 3$a$-LCO$_{2.67}$, the oxygen comes out of the capping layer directly in a short way and relative short time, whereas the oxygen migrates a long way horizontally to the edges of films to get out of the films resulting in the 3$c$-LCO$_{2.67}$. This process takes a relative long time. We believe that the anisotropic oxygen-ion migration under strain would exhibit the different ionic conducting speed and efficiency.

Next, we will discuss how the oxygen-ion migration proceeds in LCO$_{2.67}$ during the orientation switching of OVC. Two possible reaction processes at the atomic-scale can be deduced based on the atomic structures of domain boundaries (Figures 4c and 4d). When a 3$a$-LCO$_{2.67}$ is tensile-strained, the O1 shared by the CoO$_6$ octahedron and CoO$_4$ tetrahedron preferentially moves to a vacant site Vo1 in the neighbored CoO$_4$ tetrahedron along the polyhedral edge with a migration barrier of ~0.95 eV, then followed by multiple steps of ion migration along the O1-Vo1-Vo2 and O2-O1-Vo1 pathways. Finally, the oxygen ions migrate to a stable location and complete exchanges between the CoO$_6$ octahedron and CoO$_4$ tetrahedron as indicated by the blue double arrows in Figure 4c. Consequently, the orientation switching of OVC can be elucidated by cooperatively multiple octahedra-tetrahedra exchanges, as indicated by the schematics in Figure S6. In comparison, when a compressive strain is applied to 3$c$-LCO$_{2.67}$, the migration barrier (~1.03 eV) is slightly larger than that of former case. This fact well-explains that a higher $V_{\ddot{o}}$ extraction temperature is needed for compressively strained LCO$_x$ films. The final $V_{\ddot{o}}$ pattern in n-LCO$_{2.5}$ films is in sharp contrast to that of BM-LCO$_{2.5}$ on SrTiO$_3$ substrate. Interestingly, the oxygen-ion migration can be interpreted in an easier way based on the uniform CoO$_5$ configuration, as shown in Figure 4d. The migration can be accompanied just by a successive O1-Vo1 and O2-O1 hopping, leading to a simultaneous transition from tetrahedron/octahedron into the CoO$_5$ square pyramids. The neighboring octahedron would transform into CoO$_5$ square pyramids by losing one O3 ion, as indicated by the green arrow. As a result, the OVC switches towards diagonal $V_{\ddot{o}}$ stripes can be elucidated by conserved octahedra-tetrahedra evolution and partial oxygen-ion loss, as summarized by the schematics in Figure S8.

Based on the distinct OVC states and the orientation switching behaviors observed in $LCO_x$ films grown on $SrTiO_3$ and $LaAlO_3$ substates, we draw a phase diagram of OVC in strained $LCO_{3-x}$ films is established as a function of misfit strain and oxygen content in Figure 4e. Epitaxial strain determines the initial orientation of $V_{ö}$ stripes, as evidenced by the vertical stripes on the tensile strain states and the horizontal stripes on the compressive strain states. When the $LCO_x$ films are tensile-strained, the lattice structure firstly transits from $LCO_3$ into the $3a$-$LCO_{2.67}$, then further transforms into the $3c$-$LCO_{2.67}$ state. Apparently, the OVC orientation concomitant switches during the topotactic structural transition. The OVC orientation maintains when the LCOx films transit into the $2c$-$LCO_{2.5}$. We believe that the OVC orientation switching in the tensile-strained $LCO_x$ films is attributed to the strain accommodation between the final state, i.e., $2c$-$LCO_{2.5}$, and STO substrates., yielding to a +2.5% tensile strain. Of most interest, on the compressive strain regime, a new $LCO_{2.5}$ phase emerged when a large number of oxygen-ions is extracted from the $3c$-$LCO_{2.67}$ phase, forming a completely new $V_{ö}$ ordering configuration. The quantitative analysis on the $V_{ö}$-induced chemical expansion is favorable for uncovering the mechanism of OVC orientation switching. As increasing the density of $V_{ö}$ stripes, the large La-La distance $d_{Vo}$ (4.39 ± 0.05Å) in the $V_{ö}$-ordered $CoO_4$ tetrahedral layers and the shortest one $d_O$ (3.66 ± 0.05 Å) in $CoO_6$ octahedral layers appears in $3a$-$LCO_{2.67}$ due to the lattice constant confinement from $SrTiO_3$ substrates. The highly compressively strained $d_O$ makes the structure of $3a$-$LCO_{2.67}$ unstable for accepting more $V_{ö}$ and tends to switch towards the $3c$-$LCO_{2.67}$ with a relaxed $d'_O$ (3.76 ± 0.07 Å) and $d'_{Vo}$ (4.33 ± 0.05Å) along $c$ axis. In the $3c$-$LCO_{2.67}$ on LAO substrate case, the $d_{Vo}$ (3.75 ± 0.10 Å) and $d_O$ (4.47 ± 0.12 Å) evolute to be 3.80 ± 0.11 Å and 4.57 ± 0.13 Å, respectively. We summarize the calculated atomic distances in Table S1. Finally, we define the $d_{Vo}/d_O$ ratio as $V_{ö}$ tolerance factor, which determines the OVC configuration in an epitaxial thin film. The $d_{Vo}/d_O$ of ~ 1.2 is obtained in the $3a$-$LaCoO_{2.67}$ and $3c$-$LaCoO_{2.67}$ before the orientation switching of OVC. We believe the calculation of this value will serve as a criterion to evaluate the orientation stability of OVC in the oxygen-deficient perovskite oxide thin films with similar crystal structure.

In summary, we report the strain-mediated distinct oxygen-ion migration pathways in ultrathin cobaltite thin films using *in-situ* atomic-resolved STEM imaging. The evolution of OVCs is directly visualized as a function of annealing temperature and annealing time. We reveal the dynamic process of OVCs propagation and orientation switching in $LCO_x$ films under both tensile and compressive strain. Given a broad interest in the oxygen-ion conductors, we construct a phase diagram of OVC mediated by strain and oxygen content and provide an OVC stability criteria with a conception of tolerance factor. These findings suggest that the dedicated balance between chemical expansion and epitaxial strain dominate the formation and orientation of OVC, providing a practical guide for engineering targeted OVC configuration in oxygen-deficient functional oxide films. Given the abundance of perovskite oxides with ordered OVC, our work offers an exciting opportunity to design pre-determined OVC, especially the vortex-like topological ionic conductive channels, which manifest themselves to engineer unique phases and functionalities in energy materials.

## AUTHOR INFORMATION

### Corresponding Author

*Er-Jia Guo-ejguo@iphy.ac.cn

*Lin Gu-l.gu@iphy.ac.cn

### Author Contributions

† These authors contribute equally to this work. These samples were grown and processed by Q.J., S.L., and S.R.C. under the guidance of E.J.G.; TEM lamellas were fabricated with FIB milling by F.Q.M; TEM experiments were performed by Q.H.Z., F.Q.M and X.Y.L, analyzed by Q.H.Z., T.T.S and D.S; The first-principles calculations were performed by A.G., E.J.G.

and L.G. initiated and supervised the work. Q.H.Z. and E.J.G. wrote the manuscript with valuable inputs from L.G. and D.S. All authors participated in revising the manuscript.

ACKNOWLEDGMENT

This work was supported by the National Key Basic Research Program of China (Grant No. 2020YFA0309100 and No. 2019YFA0308500), the Beijing Natural Science Foundation (Z190010 and 2202060), the Strategic Priority Research Program of Chinese Academy of Sciences (Grant Nos. XDB07030200 and XDB33030200), Key research projects of Frontier Science of Chinese Academy of Sciences (QYZDB-SSW-JSC035), the National Natural Science Foundation of China (Grant Nos. 51672307, 51991344, 11974390, 52025025, 52072400) and the Beijing Nova Program of Science and Technology (Grant No. Z191100001119112).

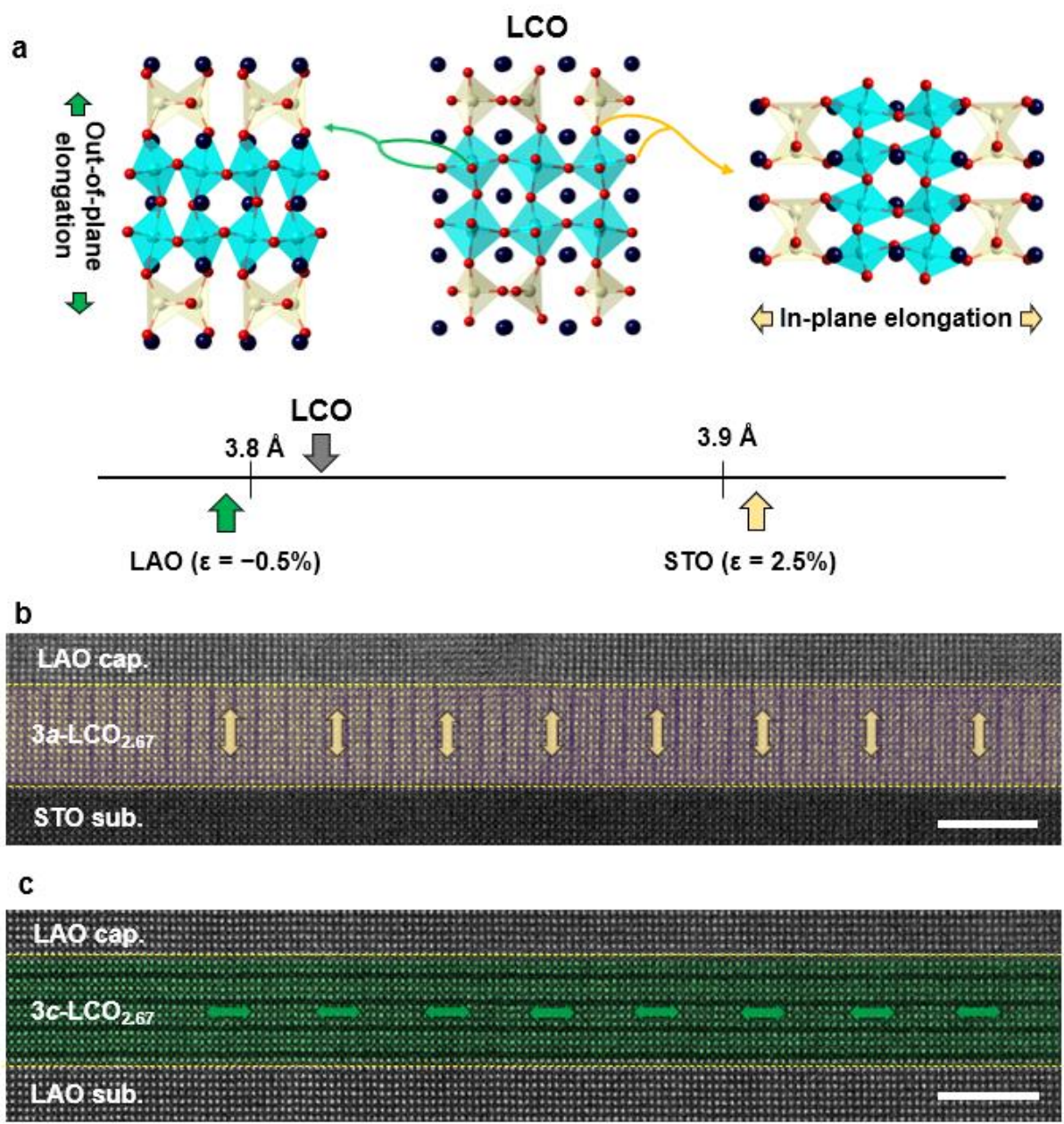

**Figure 1. Direct observations of OVCs on strained LCO$_x$ films.** (**a**) Schematics of OVCs evolution on the LaAlO$_3$ (LAO) and SrTiO$_3$ (STO) substrates. The LCO$_x$ films will form alternated tetrahedron [CoO$_4$, yellow] and octahedron [CoO$_6$, blue] along the in-plane (out-of-plane) direction under tensile (compressive) strain. (**b**) and (**c**) HAADF images of LCO$_{2.67}$ films on SrTiO$_3$ and LaAlO$_3$ substrates, respectively. 3$a$-LCO$_{2.67}$ (3$c$-LCO$_{2.67}$) represent the domain orientation aligned out-of-plane (in-plane) direction.

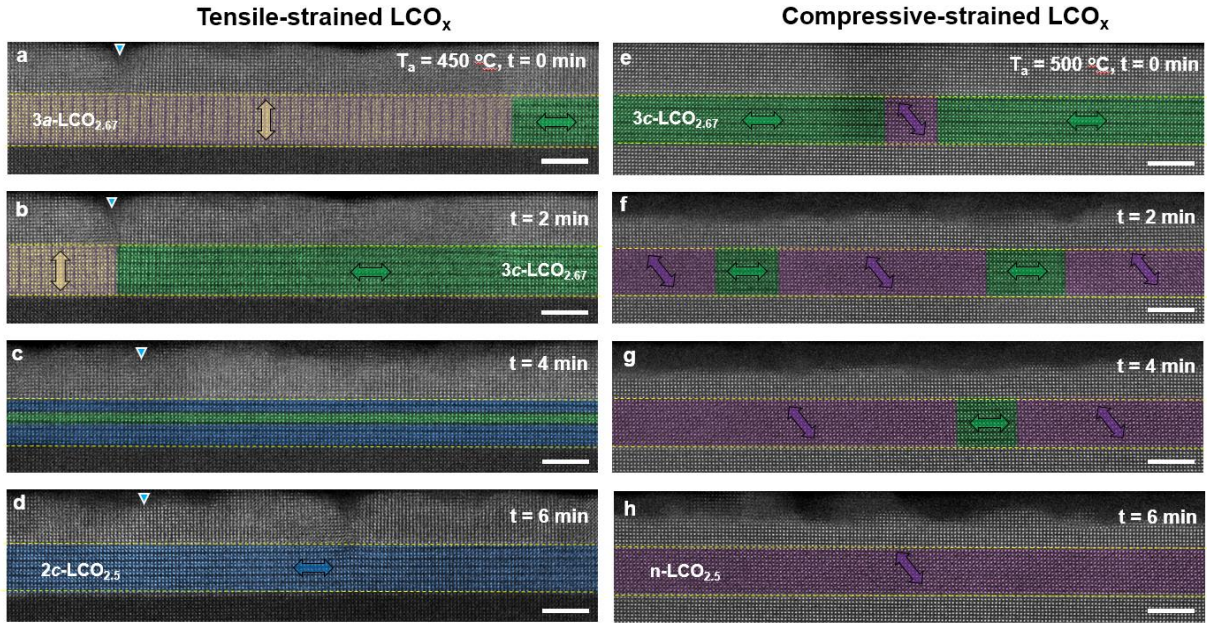

**Figure 2. Structural evolution of OVCs in strained LCO$_x$ ultrathin films.** Both specimens were *in-situ* heated up from room-temperature to annealing temperature $T_a$. (**a**)-(**d**) HAADF images of a tensile-strained LCO$_x$ film taken after annealing at $T_a$ = 450 ºC for 0-, 2-, 4-, and 6-minutes waiting time. The 3$c$-LCO$_{2.67}$ and 3$a$-LCO$_{2.67}$ domains are marked in green and yellow, respectively. After 4 minutes, the LCO$_x$ film transforms into the brownmillerite 2$c$-LCO$_{2.5}$ partially, then changes into 2$c$-LCO$_{2.5}$ (blue domains) entirely after 6 minutes. The white triangle indicates the positions of the edge dislocation. (**e**)-(**h**) HAADF images of a compressively strained LCO$_x$ film taken consequently after annealing at $T_a$ = 500 ºC for 0-, 2-, 4-, and 6-minutes waiting time. The 3$c$-LCO$_{2.67}$ transforms into a new LCO$_{2.5}$ (n-LCO$_{2.5}$) phase (purple domains). With increasing the annealing time, all n-LCO$_{2.5}$ phase merges. The white scale bar is 5 nm. Yellow dashed lines indicate the interface positions.

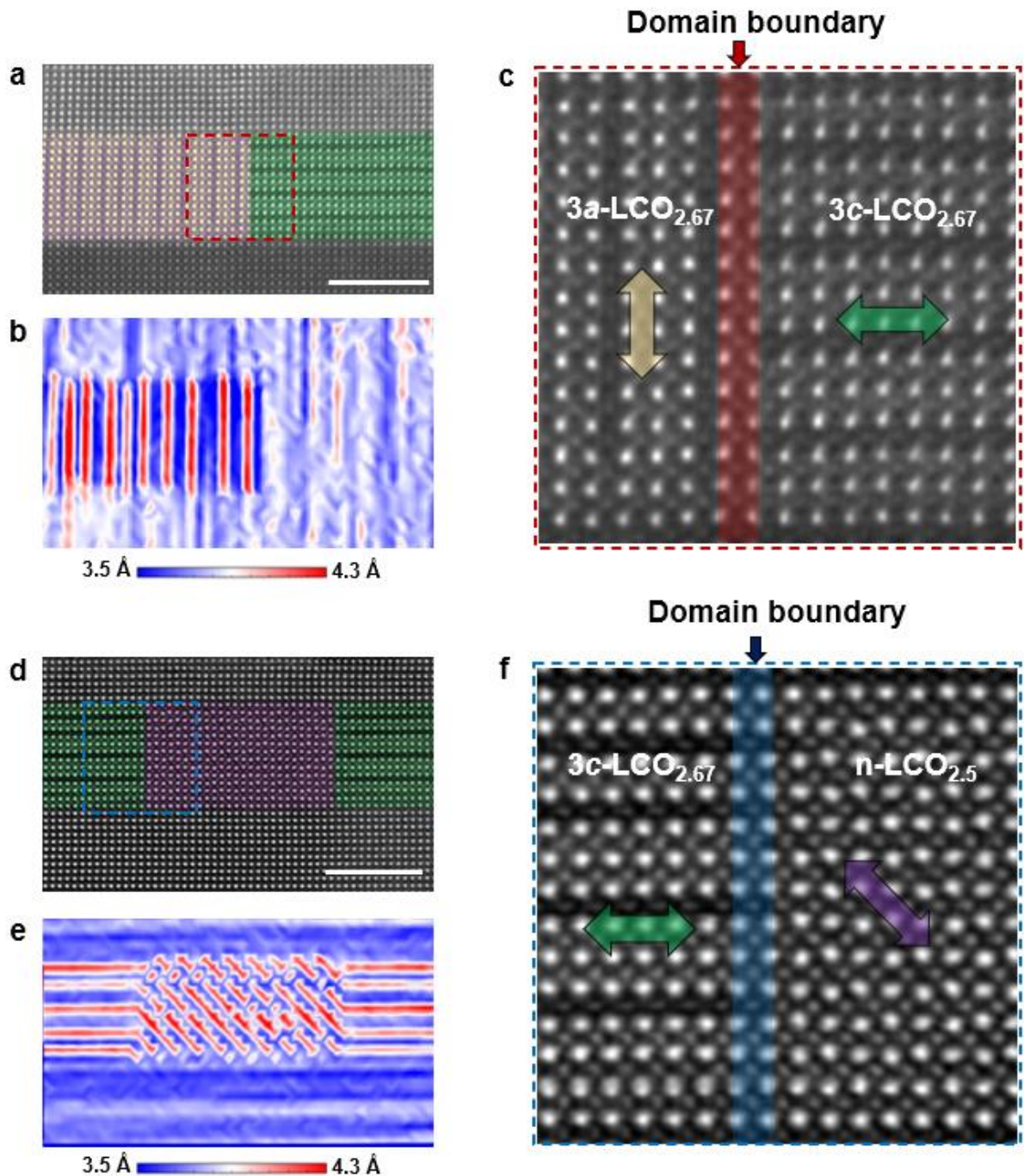

**Figure 3. Microscopic view and strain distribution of domain boundaries in LCO$_x$ ultrathin films.** HAADF images of domain boundaries (**a**) between 3$a$-LCO$_{2.67}$ and 3$c$-LCO$_{2.67}$ in a tensile-strained film and (**d**) 3$c$-LCO$_{2.67}$ and n-LCO$_{2.5}$ in a compressive strained film. High-magnified HAADF images from red and blue squares are shown in (**c**) and (**f**), where boundaries of OVC switching are indicated by red and blue colors, respectively. The in-plane and out-of-plane lattice strain mapping corresponding to (**a**) and (**d**) are shown in (**b**) and (**e**), respectively.

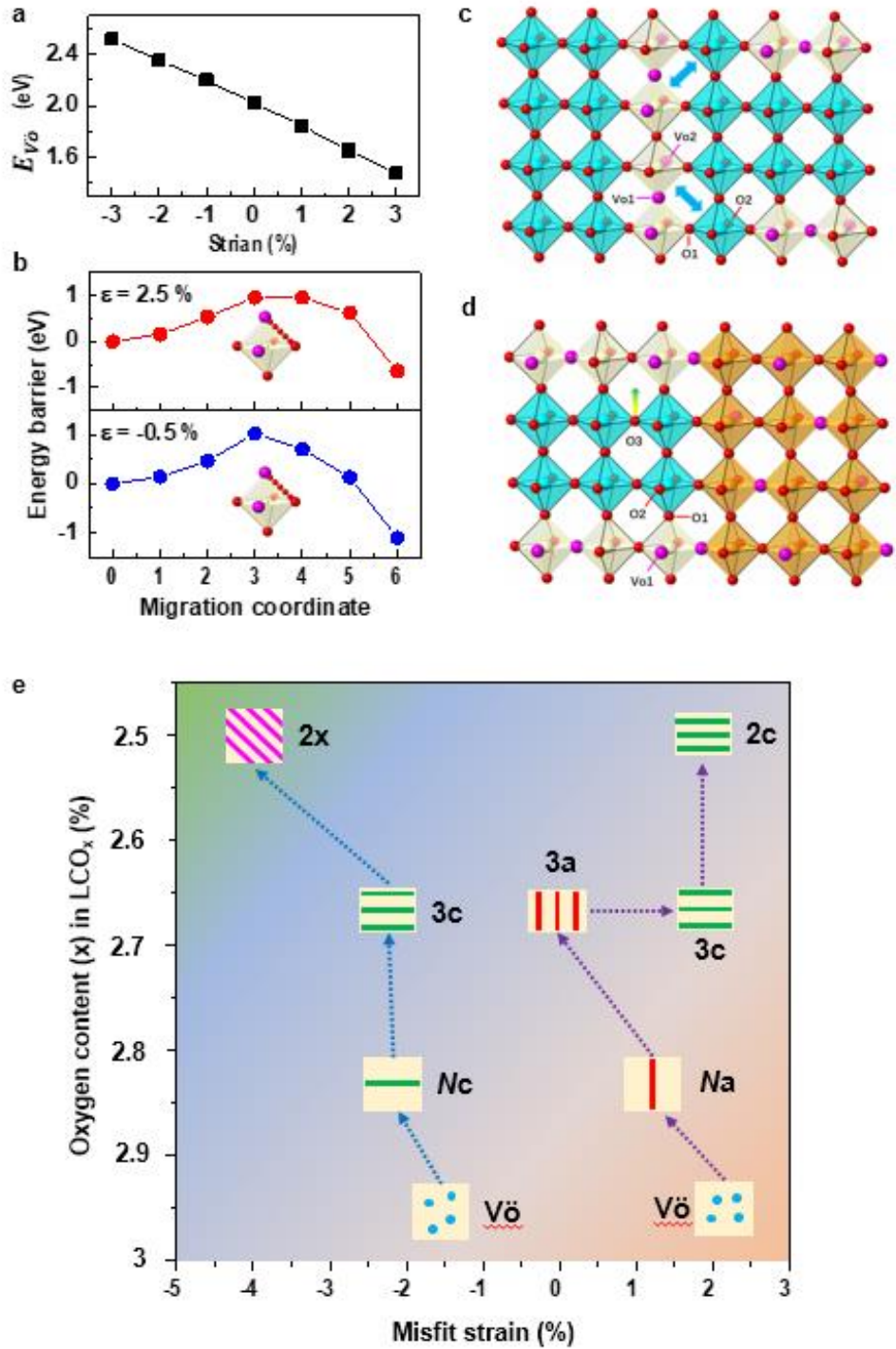

**Figure 4. Strain dependent oxygen migration and phase diagram of OVCs in cobaltites.**
(**a**) Oxygen vacancy formation energy ($E_{V_{\ddot{o}}}$) in LaCoO$_3$ film mediated by substrate strain. (**b**) The migration barrier in the CoO$_4$ tetrahedra under tensile (upper panel) and compressive strain (lower panel). Oxygen-ion migration pathway when a LCO$_x$ film under (**c**) tensile and (**d**) compressive strain. Red and purple spheres represent oxygen-ions and oxygen vacancies, respectively. (**e**) Phase diagram of OVCs. The oxygen content (x) is plotted as a function of misfit strain, which is caused by $V_{\ddot{o}}$-induced chemical expansion. $N$a and $N$c are two intermedia states of LCO$_x$ films during the phase transition under tensile and compressive strains, respectively. Note that the misfit strain and corresponding oxygen content are calculated directly from the STEM measurements.

**TOC graphics**

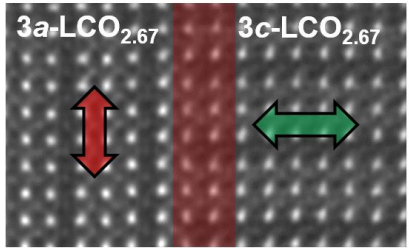
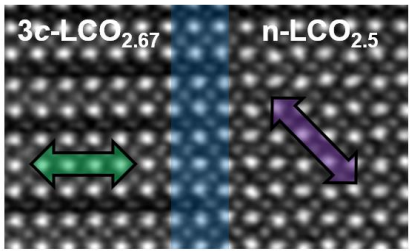
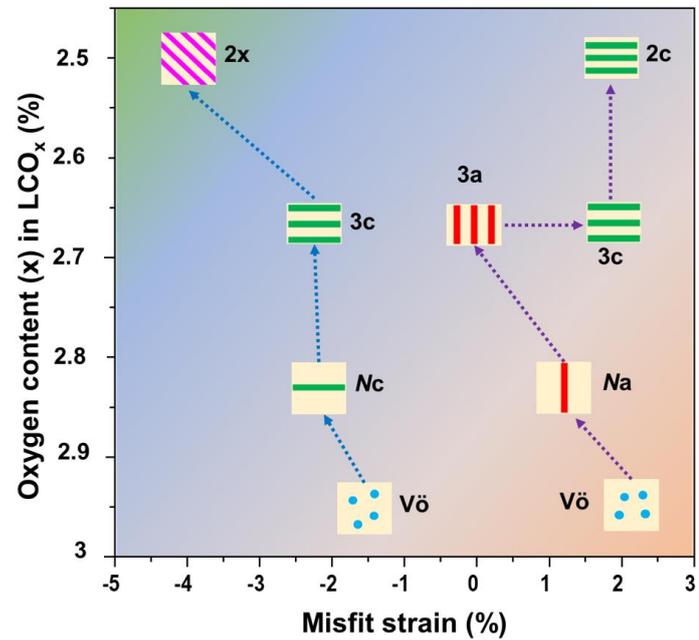

**Dynamic anisotropic oxygen-ion migration in cobaltite thin films** is observed at atomic-scale via STEM. A phase diagram of orientation of oxygen vacancy (OVC) mediated by strain and oxygen content, and an OVC stability criteria are constructed, providing a practical guide for engineering targeted OVC configuration in oxygen-deficient functional oxide films.